\documentclass[10pt,twocolumn]{article}

\usepackage{color}
\usepackage{authblk}
\usepackage{fullpage}
\usepackage{graphicx}
\usepackage{abstract}
\usepackage[numbers, square, comma]{natbib}
  
\begin{document}
\title{Correlation between Quantum Conductance and Atomic Arrangement of Silver Atomic-Size Nanowires}

\author[1,2]{M. J. Lagos}
\author[1,*]{P. A. S. Autreto}
\author[1]{D. S. Galv\~ao}
\author[1]{D. Ugarte}

\date{\vspace{-5ex}}

\affil[1]{Instituto de F\'{\i}sica ``Gleb Wataghin", Universidade Estadual de Campinas, 13083-970 Campinas SP, Brazil}
\affil[2]{Laborat\'orio Nacional de Luz S\'incrotron-LNLS, CP 6192, CEP 13083-970, Campinas–SP, Brazil}
\affil[*]{To whom correspondence should be addressed. Email: autretos@ifi.unicamp.br}

\twocolumn[
\maketitle

\begin{onecolabstract}
We have studied the effect of thermal effects on the structural and transport response of Ag atomic-size nanowires generated by mechanical elongation. Our study involves both time-resolved atomic resolution transmission electron microscopy imaging and quantum conductance measurement using an ultra-high-vacuum mechanically controllable break junction. We have observed drastic changes in conductance and structural properties of Ag nanowires generated at different temperatures ($150$ and 300 K). By combining electron microscopy images, electronic transport measurements and quantum transport calculations, we have been able to obtain a consistent correlation between the conductance and structural properties of Ag NWs.  In particular, our study has revealed the formation of metastable rectangular rod-like Ag wire (3/3) along the (001) crystallographic direction, whose formation is enhanced.  These results illustrate the high complexity of analyzing structural and quantum conductance behaviour of metal atomic-size wires; also, they reveal that it is extremely difficult to compare NW conductance experiments performed at different temperatures due to the fundamental modifications of the mechanical behavior.  
\end{onecolabstract} 
]

\date{\vspace{-5ex}}

\section{INTRODUCTION}

 The intense work of the nanoscience and nanotechnology community has increased the capabilities of researchers to produce, in a surprising controlled way, new materials and objects at nanometric scale. As a result, novel physical and chemical behaviours are frequently reported opening new opportunities for devices and practical applications. For example, at present, electronics systems play an essential role in our life and the search to improve performance at low energy cost, suggests that systems based on molecules as active parts must be studied in detail \cite{reed1997conductance}. These future devices will require a precise knowledge of the physical properties of atomic-size contacts and tiny nanowires/ interconnects. 

We must note that the mechanical and electrical properties of nanoscale metal wires (NWs) have been studied both theoretically and experimentally by different groups \cite{agrait2003quantum}. Atomic-size metal junctions can be easily generated using a simple procedure. Initially, we put into contact two clean metal surfaces and then we carefully separate them apart.  The formed metal junction is stretched until rupture, when atomic-size contacts are spontaneously generated \cite{agrait2003quantum} by the elongation induced necking \cite{callister1999d}. Applying a voltage across the junction, it is possible to assess the electric transport response during the deformation process. For nanoscale junctions, the conductance curve exhibits a well defined behaviour as a function of time (hereafter called conductance curve, CC), containing typically flat plateaus separated by abrupt jumps. These discrete conductance values are approximately integer multiples of the quantum of conductance $G_0 = \frac{2e^{2}}{h}$ (where $e$ is the electron charge and $h$ is Planck’s constant) \cite{agrait2003quantum}. This method seems to be very simple and straightforward; however each CC is associated with a different NW and probably with a different atomic arrangement. Although all CCs show plateaus and jumps, the curve profile indicates a different evolution (in fact representing the structural evolution) \cite{rego2003role}. In these terms, a statistical analysis is frequently used to analyze an ensemble of CCs, where each curve is represented by a histogram of conductance occurrence (a plateau appears as a peak). Then, the so called global histogram (GH) is constructed by a linear addition of all individual curve histograms. The GH describes the average tendency of the NW conductance measurements \cite{reed1997conductance,rego2003role,bettini2005real}, where strong peaks are related to more frequent (and probably more stable) atomic structures. 

The generation of atomic-size metal wires by the mechanical stretching has allowed the study of a wide range of metals \cite{agrait2003quantum}, including for example, Au \cite{costa1995nanowire, brandbyge1995quantized, muller1992conductance, sirvent1996stm, rodrigues2000signature, scheer1998signature}, Cu \cite{ludoph2000conductance, gonzalez2004indication}, Pt \cite{rodrigues2003evidence,rodrigues2007size}, Pd \cite{scheer1998signature, rodrigues2007size}, Al \cite{scheer1998signature}, alkaline metals \cite{yanson2000supershell}, metal alloys \cite{enomoto2002quantized,fujii2004alloying}, etc. Nevertheless, the comparison of data obtained from different experimental conditions (temperature or vacuum level) has produced some discrepancies and controversial interpretations of GH peaks \cite{costa1995nanowire, ludoph2000conductance}. In particular, the mechanical elongation induces the formation of a sequence of more favorable atomic structures during the wire thinning. Due to the dominant role of surface energy in this size regime, several anomalous wire structures have already been reported to been formed during the thinning of very tiny wires, as for example helical \cite{kondo2000synthesis} and tubular metal \cite{lagos2009observation} wires.  

The direct and simultaneous measurement of atomic arrangement and electrical transport represents quite a challenging task, achieved only in rare cases \cite{kizuka1997cross, ohnishi1998quantized, kizuka2008atomic, kurui2009conductance}. Recently, a few studies have compared the structural and/or conductance behavior of gold NWs generated at different temperatures \cite{lagos2010temperature, lagos2011temperature,lagos2011mechanical} and significant modifications were observed. This is in fact expected, because it is obvious that temperature should play a major role in the NW mechanical behavior, defect generation/annihilation, etc. \cite{bratkovsky1995conditions}. Although significant progress has been achieved on the understanding of metal NW properties, the majority of the available experimental data is related to gold-based wires due to their easy sample preparation and manipulation. In order to obtain more broad-ranging insights, we should study other model systems, as for example another metal with different surface properties (in order to induce different atomic structure during elongation). Silver represents an excellent candidate, because it is also a noble metal with identical crystal structure (face centered cubic, fcc) and almost the same lattice parameter as gold. In contrast, silver facets with the minimal energy facets are the (100) instead of the (111) ones in Au \cite{foiles1986embedded}. It must be emphasized that the higher reactivity of silver certainly renders much more difficult the studies and analysis. Here, we present an experimental and theoretical study of thermal effects on electronic transport and structural behavior of atomic-size Ag wires generated by mechanical elongation. We have used dynamical atomic resolution transmission electron microscopy (HRTEM), conductance measurements, and quantum transport calculations, to consistently understand the origin of the observed remarkable modifications of GH from experiments realized at $\sim 150$ and $\sim 300K$. 

\section{METHOD}

 As mentioned above, in this work we have used different approaches to carry out an extensive study of the transport properties of silver NW´s generated at $300$ and $150K$. Firstly, we have performed electrical conductance measurements at both temperature conditions. Also, the atomic arrangement of atomic-scale Ag wires was analyzed using in situ transmission electron microscopy observations. Finally, we have used theoretical calculations of quantum conductance for different NW structures to interpret the most significant peaks in GH of Ag NWs.

Electrical conductance measurements were performed in a dedicated instrument: mechanically controllable break of junctions (MCBJ), which operates in ultra high vacuum conditions (UHV, $< 10-10$ mbar). In this technique, a very thin silver wire ($75 \mu m$ in diameter and $99.99 \%$ pure) is notched and attached to a flexible substrate (Cu-Be alloy), then the substrate is slowly bent by means of a piezoelectric-based linear mechanism, producing the rupture of the wire at the notch. In this way, two clean surfaces are generated, as the breaking process is performed in situ in UHV environment. The formation and rupture of NWs is achieved by putting into contact the generated surfaces and subsequently retracting them. The UHV-MCBJ allowing the measurement at $300$ and $150K$ has been thoroughly described elsewhere \cite{rodrigues2008low}. The measurement electronics is basically composed of a home-made voltage source and a current-voltage converter coupled to an eight-bit digital oscilloscope (Tektronic TDS540C). The acquisition system input impedance and time response were optimized to perform conductance measurements in the ($0 - 5.5 G_{0}$) range with a relative error \cite{rodrigues2008low} of $\Delta G/G$ $\sim 10^{-4}$. It must be emphasized that before the start of the conductance measurements, degassing/cleaning process is performed in order to reduce contaminant presence in the Ag NWs during conductance measurements. A Joule heating of the filament is realized by passing a $1-2$ A current for few weeks. Initially, the wire degassing generates a pressure increase inside the UHV chamber; finally optimal vacuum conditions are afterwards recovered (period of $15-20$ days). 

The atomic structure of the Ag wires was directly observed using a HRTEM microscope (JEM-3010 URP, $0.17$ nm point resolution) equipped with a standard and liquid nitrogen (LN) sample holder (Gatan 613-DH, sample temperature attains $\sim 150K$ \cite{lagos2011temperature, oshima2003helical}. The wires were generated in situ following the procedure proposed by Kondo and Takayanagi \cite{kondo1997gold}, where nanometric holes are created in a self-supported polycrystalline thin metal film ($\sim 40-50$ nm in thickness) by focusing the HRTEM electron beam. The metal nanobridges formed between neighboring holes, spontaneously thin and break, generating atomic-size Ag NWs. The dynamical evolution of the Ag NW was recorded using a high sensitivity TV camera (Gatan 622SC, $30$ frames/s) coupled to a standard DVD recorder. The atomically resolved images were acquired close to Scherzer defocus with a current density of $\sim 10-30$ A $cm^{-2}$. The HRTEM images presented here are snapshots extracted from DVD recordings; a detailed description of the HRTEM procedures to study metal nanowires has already been presented elsewhere \cite{bettini2005real}. 

Finally, in order to establish a consistent correlation between the conductance and structural information, we have carried out theoretical conductance calculations using the procedures introduced by Emberly and Kirczenow \cite{emberly1999electron} based on Extended Huckel Theory (EHT). In this methodology, the electronic quantum transport is simulated within the Landauer scattering formalism using the molecular orbitals (MO) obtained via EHT. This approach has been already applied with success in studies of metal NW’s \cite{rego2003role, gonzalez2004indication, lagos2010temperature, rodrigues2002quantum}. In this study, the NWs were coupled at both sides to two semi-infinite leads; MOs were calculated taking into account s, p and d Ag orbitals, as well as, overlap and energy matrix elements extending beyond first-neighbor atoms. It should be stressed that the wire length is an important issue in the calculations. If the wire is not long enough to preclude an artificial overlap among the orbitals from the apexes and the wires, this could produce an unrealistic increase in the calculated conductance values. Once the wire is long enough to preclude these artificial overlaps, the results are independent on the wire length. We have tested wires of different lengths in order to guarantee that these needed requirements are satisfied.The experimental results are contrasted with the conductance values at Fermi energy ($E_f$).  

\section{RESULTS}

\begin{figure}[t]
\includegraphics[scale=0.38]{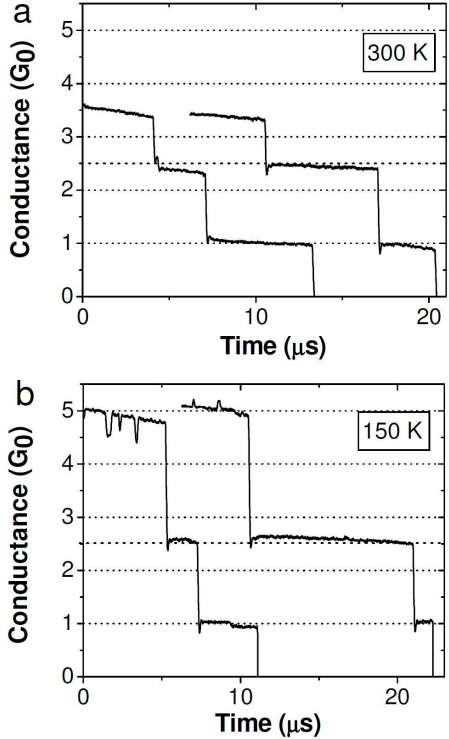}
\caption{Typical conductance curves obtained during the elongation and rupture of Ag nanowires with plateaus spaced by abrupt jumps acquired (by means of a UHV-MCBJ) at (a) 300 K and (b - c) 150 K.}
\label{fig1}
\end{figure}

Figure \ref{fig1}(a) and (b) show typical quantum conductance curves of silver NWs acquired at $300K$ and $150K$, respectively. Note that these curves exhibit approximately flat plateaus spaced by abrupt jumps. These curves profiles have been selected because they appear frequently during the wire thinning in the electric transport experiments. At room T, we have noted that about $\sim 5-10\%$ of the curves show a series of three plateaus, one at $\sim 3.5$, a second plateau just below $2.5$ and, a final one at $1.0\: G_0$ (see Figure \ref{fig1}(a)). This plateau sequence is rarely observed at $150K$; in contrast CCs with plateaus at $\sim 5.0$, slightly above $2.5$ and $1.0\: G_0$, are observed with similar percentage. The results obtained at different temperatures show unambiguously that thermal energy induces substantial variations on the NW conductance behavior, what in fact is just a conductance representation of the structural evolution during wire stretching.

\begin{figure}[ht]
\includegraphics[scale=0.35]{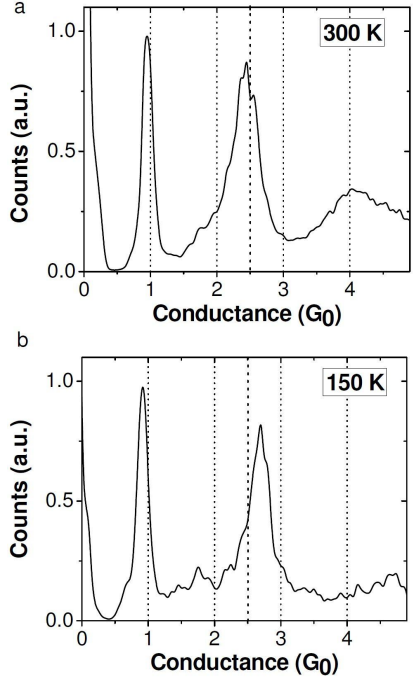}
\caption{Conductance global histograms obtained from 500 curves measured at $300K$ (a) and $150K$ (b). Both histograms have been normalized in relation to the peak at $1 G_0$; dotted lines indicate conductance which are integer multiples of $G_0$ and the dashed line indicate a conductance value of 2.5 G0. Note that prominent peaks appear at $\sim 2.4 G_0$ in (a), seem to be upwards-shifted to $\sim 2.7 G_0$ in (b).}
\label{fig2}
\end{figure}

Figure \ref{fig2} displays the derived global histograms using the conductance curves measured at $300K$ and $150K$. The main feature of both GHs is the peak located at $\sim 1 G_0$, which is associated with one-atom-wide contacts and suspended linear atomic chains (LACs) \cite{agrait2003quantum, ohnishi1998quantized}. A simple visual inspection shows that the $300K$ GH exhibits also two important peaks located at $\sim 2.4\: G_{0}$ and $\sim 4\: G_{0}$, in agreement with a previous report by Rodrigues et al. \cite{rodrigues2002quantum}.  As expected, the significant changes in CC profiles shown in Fig. \ref{fig1} are reflected in the important modifications in the $300$ and $150K$ GHs. For example, note that the prominent narrow peak, located between $2.0$ and $3.0\: G_{0}$, appears somewhat upwards shifted at low temperature, with a maximum located at $\sim 2.7\: G_0$. In addition, the room T wide peak present at $\sim  4\: G_0$ has disappeared at $150K$, while a small peak located at $\sim 1.8\: G_0$ has became easily discernible. 

\begin{figure}[!ht]
\centering
\includegraphics[scale=0.38]{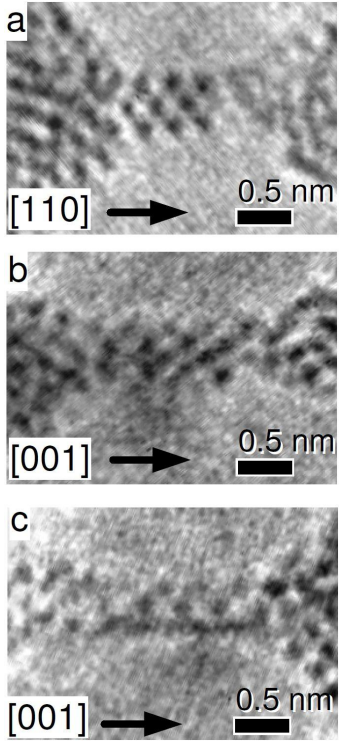}
\caption{HRTEM images of Ag NWs being stretched at $300$ (a) and $150K$ (b,c). At room temperature Ag rod-like wires (a) are crystalline, defect-free and their axis follows the (110) direction; in contrast at low temperature (a) pillar-like wires show an axis along the (001) direction. The nano-rod displayed in (c) shows the bamboo-like contrast expected for the Ag tubular structures reported by Lagos et al. \cite{lagos2009observation}. The images represent snapshots extracted from video recordings; atomic positions appear dark.}
\label{fig3}
\end{figure}

Considering the atomic structure of Ag NWs generated at different temperatures, we have also observed major differences by means of in situ HRTEM experiments. During the elongation of silver junctions at $300K$, rod-like wires oriented along (110) directions are frequently observed (hereafter noted as (110) NW, see Fig. \ref{fig3}(a)); in addition, (001) silver rods can also be observed, but they mostly break abruptly at room temperature \cite{bettini2005real, rodrigues2002quantum}. This contrasts with low temperature observations where rod-like NWs along (100) showed higher occurrence rate than (110) ones (Fig. \ref{fig3}(b)-(c)). Finally, we have also observed the formation of one-lattice-parameter-wide (100) silver rods, which can show a bamboo-like contrast  both at room- and low-temperature experiments (see Fig. 3(c)); Lagos et al. \cite{lagos2009observation, autreto2011intrinsic} have shown that this (100) wires display an anomalous tubular atomic arrangement. 

\section{DISCUSSION}

The conductance results described above are quite different from those ones obtained for gold NWs, where the experimentally observed peaks are located at $1 G_0$, $\sim 1.8\: G_0$ and $\sim 2.8\: G_0$ \cite{lagos2010temperature, lagos2007structural}. The formation of GH peaks located at different positions is expected due to the fact that gold and silver show different surface properties and, in consequence, we must expect the formation of NWs with different morphologies \cite{rodrigues2005structural}.

We must keep in mind that a particular GH peak reflects statistically relevant conductance plateaus (which are associated with the most frequently occurring atomic arrangements). These plateaus represent just one intermediate stage in a conductance curve; or in structural terms, an intermediate atomic arrangement in a series of more favorable structures generated during the stretching. It must be emphasized that the consistent explanation of the physical origin of a conductance plateau, requires understanding of the whole conductance curve, i.e., all plateaus in the curve (and the associated structures) sequentially appearing until the wire rupture. For example, it has already been demonstrated that at room temperature the Ag GH  peak at $\sim 2.4\: G_0$ is associated with the formation of one stable (110]) rod-like Ag NW \cite{bettini2005real,rodrigues2002quantum}. In fact, a sequence of three atomic arrangements whose conductance displayed values of $\sim 3.8\: G_0$, $\sim 2.4\: G_0$ and $1\: G_0$ was proposed in order to correlate structural and electronic properties of Ag NW’s deformed at $300K$. Our experimental data (see the conductance curve in figure \ref{fig1}(a) and GH in figure \ref{fig2}(b)) is in excellent agreement with these previous reported results.

The main difference between the Ag NW conductance behavior at $300$ or $150K$ is the shift of the prominent GH peaks located around $2.4 G_0$, which moves up to  $2.7\: G_0$ (see Figure \ref{fig2}).  This upward shift of conductance measured at low temperature has somewhat puzzled us. This contrasts with previous experimental studies on Au NW’s, where it has observed that, at low temperatures, the Au histogram peaks move slightly downward due to the formation of structural defects in the wire  and leads (apexes) at $150K$ \cite{lagos2010temperature, lagos2011mechanical}.  These defects act as back-scattering agents inducing a slight decrease of quantum conductance \cite{lagos2010temperature, emberly1999electron}. In contrast, our transport experiments do not show a downward shift of GH peaks, and then another effect must be taken into account.  In these terms, we must examine if another silver NW atomic arrangement may account for the $2.7\: G_0$ peak and the related CC results. As mentioned above, rod-like NWs along (001) axes were more frequently observed in HRTEM images at $150K$; we will then analyze this family of wires in more detail below.

\begin{figure}[!ht]
\includegraphics[scale=0.38]{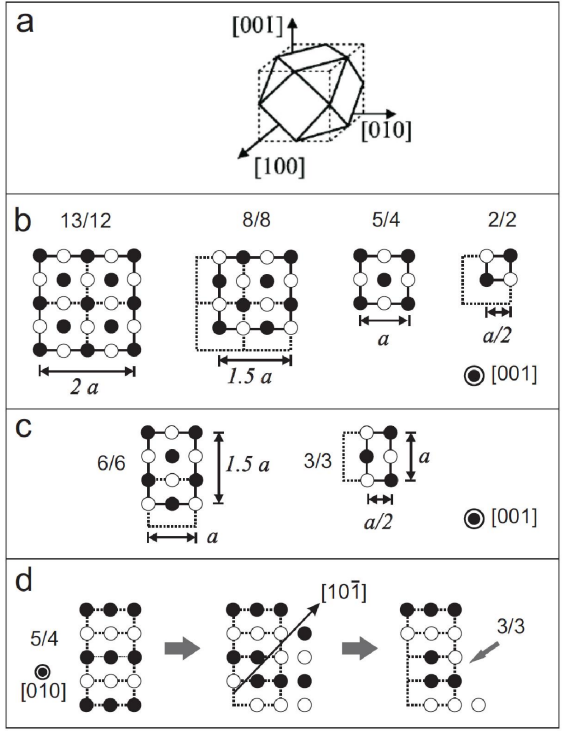}
\caption{Application of the Wulff method to determine the shape of silver NWs. (a) Expected morphology of a silver nanoparticle. (b) Cross-sectional view of NWs elongated along (001) direction with $2$, $1.5$, $1$, and $0.5$ lattice parameter ($a_0$) width; the square wires shape is generated by (100) and (010) facets. Note that (001) NWs are formed by the stacking of two (200) planes along the axis (displayed with different colors). (c) Cross-sectional view of NWs elongated along (001) direction with rectangular shape; the wire shape is also determined by (100) and (010) facets. (d) Proposed mechanism for the generation of rectangular rod-like Ag wires.}
\label{fig4}
\end{figure}

In order to propose possible three-dimensional structures of silver NW’s, the geometrical Wulff construction \cite{rodrigues2005structural} can be used. This method yields the crystal shape by predicting the relative size of the lower-energy facets of a nanocrystal; for example, it is well-known that for Au compact (111) facets are the preferred ones and, then, gold nanosystems evolve to expose mainly these low energy facets. Fig. \ref{fig4}(a) shows the expected morphology of a Ag nanoparticle \cite{emberly1999electron}, a truncated cuboctahedron with regular triangular {111} facets.  Recently, the same approach has been successfully applied to model metal nanojunctions generated by mechanical elongation \cite{rego2003role, gonzalez2004indication, rodrigues2002quantum}. We must keep in mind that, in order to get a realistic description of the NW conductance, it is mandatory to attach the NW to two pyramidal apexes at its ends. As NW apexes must also follow the same metal surface energy constrains, they were also derived according to Wulff method, as already applied to study Au, Cu, or Pt NWs \cite{rego2003role, rodrigues2000signature, gonzalez2004indication, rodrigues2003evidence}.

Figure \ref{fig4} shows the application of the Wulff approach to estimate the morphology of (001) rod-like Ag wires ($0.5$ to $2$ lattice parameter ($a_0$) in width). From Figure \ref{fig4}(a), it is possible to easily note the high relevance of the square {100} facets over {111} facets.  In these terms, surface energy minimization predicts (100) and (010) facets (which parallel to the wire axis), then the rod-like wire cross-section should be square (see Fig. \ref{fig4}(b)). The (001) NWs are formed by the stacking of two (002) planes along the axis (displayed with different colors in Fig. \ref{fig4}(b)). Hereafter, we will name them using the number of atom in each of these planes (ex. 5/4 for the $a_0$ -wide (001) NW, see Fig. \ref{fig4}(b)). On this basis, it is intuitive to think that squared-cross-section (001) NW should represent the preferred atomic arrangement to be generated during the mechanical elongation of a face centered cubic (FCC) metal such as silver \cite{lagos2009observation}. The structures displayed in Figure \ref{fig4}(b) constitute a sequence of atomic arrangements that should appear during the thinning of square Ag (100) NW (sequentially with $2$, $1.5$, $1$, and $0.5\: a_0$ in width). They must finally lead to the formation of a suspended atom chain.  

\begin{figure}[!ht]
\includegraphics[scale=0.38]{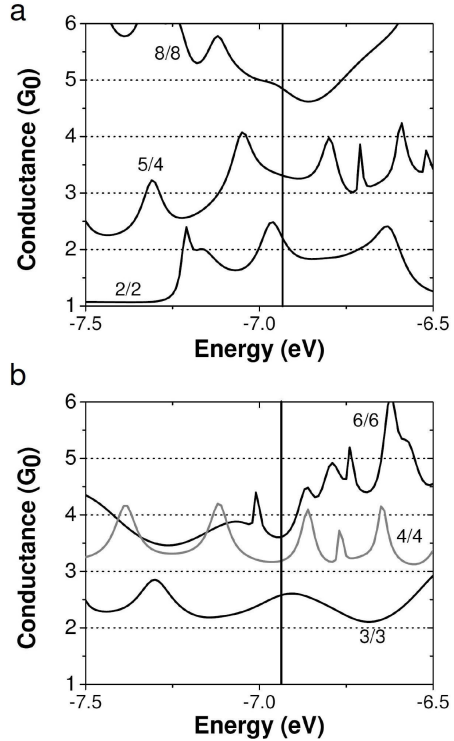}
\caption{Theoretical calculations of quantum conductance as a function of energy for silver nanowires with different morphologies. (a) conductance associated with the atomic arrangements shown in Figure \ref{fig3}(b) (8/8, 5/4, 2/2 (100) NWs). (b) conductance associated with rectangular wires (6/6 and 3/3) and, also the tubular (4A/4B) Ag nanotube.) . The conductance is plotted in units of $G_0$ and, the vertical line indicates the Fermi energy.}
\label{fig5}
\end{figure}

At this point, we must try to use the proposed atomic structures of (001) NWs to explain the quantum conductance curve behaviour at $150K$.  Figure \ref{fig5}(a) shows the calculated conductance associated with the proposed wire structures shown in Figure \ref{fig4}(b). Initially, we observe that the square (001) NWs show conductance values at Fermi energy of about $5.0$, $3.3$ and $2.1\: G_0$ for the 8/8, 5/4 and 2/2 wires respectively. These results indicate the expected sequence of conductance plateaus from (100) NWs and, finally a $1 G_0$ plateau associated with LACs (we have also performed this calculation, not shown here). The conductance curves in Figure \ref{fig1}b do show the 5 and $1\: G_0$ plateaus; also the $150K$ GH (Figure \ref{fig2}b) indicates the existence of rather small wide peak around $4.8\: G_0$. Surprisingly, no significant GH intensity or peak is present in the $3-4\: G_0$ region, somewhat suggesting the low statistical relevance of the very intuitive one-lattice-parameter 5/4 Ag rod. Also, the occurrence of the 2/2 wire does not seem to be statistically meaningful, but it might be related to the subtle peak around $1.8\: G_0$. 

In view of the fact that square (001) Ag wires based on the macroscopic fcc structure can not account for the major GH peak at $2.7\: G_0$, we must put forward another hypothesis; in this sense, HRTEM images provide very important information. Figure \ref{fig3}c shows the formation of an anomalous one-lattice parameter-wide tubular (100) Ag wire, which was recently reported by Lagos et al. \cite{lagos2009observation}. The atomic arrangement of this tubular NW, is derived from the 5/4 square rod displayed in Figure \ref{fig4}(b) and, removing the central atom of the plane containing $5$ atom. In this way, we generate a 4A/4B stacking sequence along the (100) direction \cite{lagos2009observation}. Our calculations predict that this tubular atomic arrangement should display a conductance of about $3.1-3.2\: G_0$ (Figure \ref{fig5}(b)). As mentioned above, the GH acquired at $150K$ do not show any peak in the $3-4\: G_0$ region, indicating a low occurrence of this structure in MCBJ experiments. We must keep in mind that the formation of the Ag nanotube was associated with high deformation rates \cite{lagos2009observation}, a parameter that can not be controlled/reproduced accurately in our MCBJ experiments. 

As matter stands, the experimental conductance measurements can not be correlated satisfactorily by square rod-like Ag wires (fcc or tubular). The recent paper of Lagos et al. \cite{lagos2011temperature} uncovering the existence of the anomalous tubular Ag wire, also highlights the important fact that the mechanical deformation of nanoscale systems may lead to the formation of high-symmetry metastable atomic arrangements fulfilling surface energy minimization requirements. Following these ideas and looking carefully at the square (001) NW structures in Figure \ref{fig4}(b), it becomes obvious that we can not rule out the formation of intermediate (001) wires with rectangular cross-section, when (010) and (001) facets are considered. These wires should represent local minima of surface energy minimization within Wulff criterion during the process of mechanical elongation; Figure 4 (c), show two examples of rectangular wires (marked as 6/6 and 3/3 ones). From the plastic deformation point of view, the generation of rectangular (001) NWs is not surprising; in fact the expected structural flow of a square Ag (001) rod is the glide over any one of the four possible {111} compact atomic planes. For a total (110) dislocation glide, wires with rectangular cross section should be spontaneously be generated from a square Ag rods (see schematic deformation in Fig. \ref{fig4}(d), where a 5/4 wire transforms into a 3/3 one). Our calculations indicate that the rectangular (001) NWs, 6/6 and 3/3 wires, display a conductance of about $3.9$ and $2.6\: G_0$, respectively (Figure \ref{fig5}(b)). 

Thus, combining square and rectangular wires, the calculated conductance sequence associated with a $8/8 \rightarrow  3/3 \rightarrow LAC$ ($5.0 \rightarrow 2.7\rightarrow 1\: G_0$), allows a consistent correlation between atomic arrangement and quantum conductance for whole structural evolution of (001) NWs at low temperature. This sequence is in excellent agreement with the experimental CC shown in Fig. \ref{fig1}(b), confirming a correlation between the proposed geometric models and the GH peaks. The rectangular 3/3 NW must be associated with the formation of the prominent peaks around $2.7\: G_0 $ in the 1$50K$ GH (Figure \ref{fig2}(b)). In these terms, the lowering of the temperature has lead to a drastic  modification of the preferred or energetically favoured structure during elongation or, in more general terms,  significant changes of the Ag NW structural behaviour: a) the dominant kind of wire changes from (110) wires at $300K$ to (001) ones at $150K$; b) the major peak in the $150K$ GH  ($\sim 2.7\: G_0$), which should be  associated with  the more frequent (and probably more stable) atomic structure, is accounted for a presumably metastable rectangular rod-like wire. 

In order to further elucidate this issue, we must first remind that the NWs studied here are generated by means of mechanical stretching and the conductance is measured simultaneously to the wire thinning. Then, in this work, the drastic changes in mechanical response are revealed through the conductance GHs, which in addition provide statistical averaging of possible atomic structures. As in many other physical phenomena, size effects have strong influence on the active deformation mechanisms in nanoscale materials and, also, thermal effects may induce drastic changes in mechanical response. For example, time-resolved atomic resolution transmission electron microscopy (HRTEM) has revealed that no extended defect could be observed in 1-nm-wide gold nanorods stretched at room temperature, while at $150K$, planar defects are frequently generated \cite{lagos2011temperature}. The planar defects are generated by the compact glide of the (111) planes as predicted by simulations performed by Landman et al. \cite{landman1990atomistic}. These authors have concluded that, for each temperature, the active plasticity deformation mechanisms (and the derived mechanical behavior) in gold nanowires are strongly dependent on wire size and shape. Extending these ideas to Ag wires, it would not be surprising to observe similar behavior modification in Ag NW structures with the glide over {111} planes as the main deformation mechanism. When a nanowire is allowed to generate defects (as observed in low temperature HRTEM experiments) its strength is reduced and the stress-strain curve indicates the start of plastic flow for smaller strain values. The application of tensile force along the {001} direction in a fcc crystal as Ag, provide four possible (111) glides planes, making (001) NWs much more ductile and easy to elongate that along another crystallographic directions. This may be at the origin of the higher statistical occurrence of (001) NWs in silver for stretching processes at $150K$.  Also, as gilding on (111) planes should be the deformation mechanism leading to rectangular wires (ex. 3/3); the fact that there are four family of gliding planes available for (001) NWs could explain the proposed high occurrence of rectangular 3/3 rod-like Ag wires, reinforcing the interpretation of the dominant $2.7 G_0$ GH. 

It is important to mention that we do not have at the moment a definitive explanation for the differences observed to the event statistics at 150 and 300 K. At present, the plastic behavior of nanoscale matter is far from be well understood. Recent experiments \cite{lagos2011mechanical, brinckmann2008, kim2009tensile} have revealed a quite different dynamics of deformation of different metallic nanowires at low and room temperatures. These differences are greater than what we could expect only in terms of the small differences of thermal energies associated with 150 and 300 K and are suggestive of a complex dynamics that goes beyond pure thermal effects. More studies are necessary to a better understanding of these fundamental aspects. We hope that our present work will stimulate further theoretical and experimental works along these lines.

In summary, we have observed drastic changes in conductance and structural properties of Ag nanowires generated by mechanical elongation for experiments realized at different temperatures ($150$ and $300K$). By combining atomic resolution HRTEM images, UHV-MCBJ transport data and theoretical calculation we have been able to obtain a consistent correlation between the conductance and structural properties of Ag NWs. Significant changes of the Ag NW structural behaviour with temperature have been observed. Firstly, the main peak of the GH kind is associated with (110) wires at $300K$  and changes to (001) ones at $150K$. Secondly, the major peak in the $150K$ GH  ($\sim 2.7\: G_0$), associated with  the more frequent atomic structure, is accounted for a metastable rectangular rod-like wire (3/3) along (001) direction, whose formation is enhanced due to a favored geometrical symmetry considering the main active deformation mechanism in atomic size fcc wires (i.e compact glide over (111) atomic planes).  The occurrence of this kind of rod-like wire would have been rather difficult to predict without its quantum conductance signature. These results demonstrate the high complexity of analyzing structural and quantum conductance behaviour of metal atomic-size wires generated by mechanical stretching. In particular, our study reveals that it is extremely difficult to compare of NW conductance experiments performed at different temperatures due to the crucial modifications of the mechanical behavior.

The authors are grateful to FAPESP, CNPq and, LNLS for financial support

\bibliography{bibliografia}{}
\bibliographystyle{plain}

\end{document}